\documentclass{article}

\usepackage{makeidx}         % allows index generation
\usepackage{graphicx}        % standard LaTeX graphics tool
\usepackage{multicol}        % used for the two-column index
\usepackage{cite}            % adjusts the "syntax" of the refs in the
\usepackage[bottom]{footmisc}% places footnotes at page bottom
\newcommand{\mbf}{\mathbf}
\newcommand{\mbs}{\boldsymbol}
\usepackage{amsmath}
\usepackage{wrapfig}
\makeindex             % used for the subject index
\begin{document}

\title{Strained graphene structures: from valleytronics to pressure sensing}
\author{Slavi\v{s}a P. Milovanovi\'{c}$^{1, \dagger}$ and Fran\c cois M. Peeters$^{1, \ddagger}$}
\date{ $^1$University of Antwerp, Groenenborgerlaan 171, B-2020 Antwerp \\
        $^\dagger$\texttt{slavisa.milovanovic@uantwerpen.be} \\
        $^\ddagger$\texttt{francois.peeters@uantwerpen.be} 
}

\maketitle

\begin{abstract}
Due to its strong bonds graphene can stretch up to 25$\%$ of its original size without breaking. Furthermore, mechanical deformations lead to the generation of pseudo-magnetic fields (PMF) that can exceed 300 T. The generated PMF has opposite direction for electrons originating from different valleys. We show that valley-polarized currents can be generated by local straining of multi-terminal graphene devices. The pseudo-magnetic field created by a Gaussian-like deformation allows electrons from only one valley to transmit and a current of electrons from a single valley is generated at the opposite side of the locally strained region. Furthermore, applying a pressure difference between the two sides of a graphene membrane causes it to bend/bulge resulting in a resistance change. We find that the resistance changes linearly with pressure for bubbles of small radius while the response becomes non-linear for bubbles that stretch almost to the edges of the sample. This is explained as due to the strong interference of propagating electronic modes inside the bubble. Our calculations show that high gauge factors can be obtained in this way which makes graphene a good candidate for pressure sensing.
\end{abstract}
\section{Introduction}
\label{sec:0.0}
Graphene is the first ever two-dimensional material to be synthesized in a laboratory\cite{rgr1}. Before K. Novoselov and A. Geim used mechanical exfoliation to isolate one layer of carbon atoms from a piece of graphite, it was firmly believed that two-dimensional (2D) materials can not exist in nature except as part of three dimensional structures. This belief was based on various experiments with thin films and on theoretical considerations, i.e. the Mermin-Wagner theorem \cite{r_add_4}. These experimental studies showed that the melting temperature of thin films rapidly decreases with decreasing thickness \cite{r45, r46}. Hence, it was believed that thermal fluctuations would lead to segregation and decomposition of 2D materials.

"Material that shouldn't exist" triggered a lot of attention in the scientific community. Graphene showed that the two-dimensional (2D) crystals are not only feasible but turn out to be high quality structures with unique properties and numerous potential applications. This caused an enormous interest in 2D materials and a quest for new atomic thin materials started. Dozens of them have been discovered and together with materials created by stacking different 2D materials, i.e. heterostructures, the number goes well above a hundred. It is expected that there are around 500 different two-dimensional materials \cite{r48}. 

\begin{figure}[htbp]
\begin{center}
\includegraphics[width=8cm]{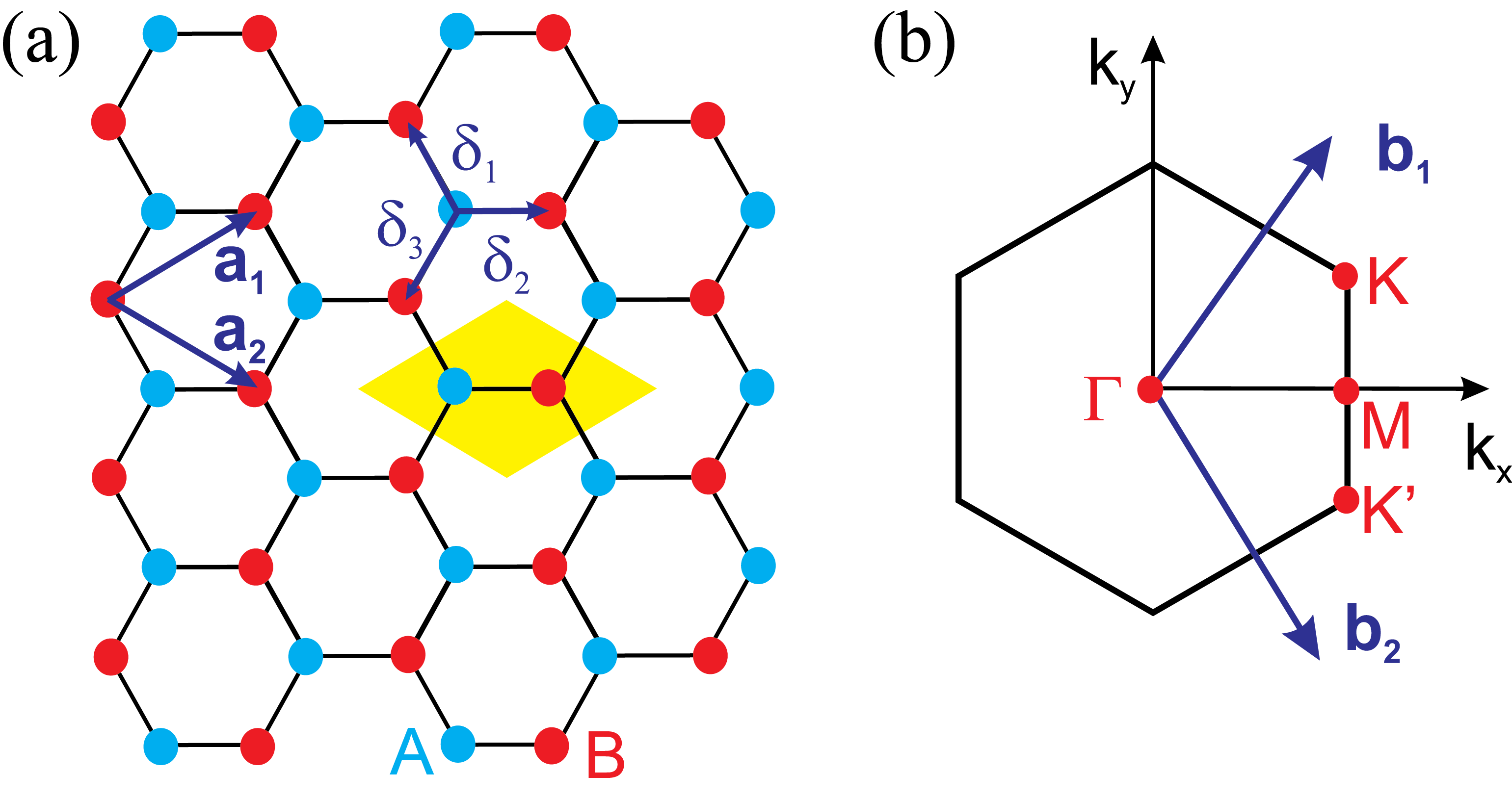}
\caption{a) Graphene lattice with basis vectors $\mbf{a_1}$ and $\mbf{a_2}$. We also show the space vectors that connect two sublattices (A and B) $\mbs{\delta_1}$, $\mbs{\delta_2}$, and $\mbs{\delta_3}$. Yellow rhombus represents the primitive unit cell. b) First Brillouin zone with reciprocal lattice vectors $\mbf{b_1}$ and $\mbf{b_2}$. Taken from Ref. \cite{rthes}.}
\label{lattice}
\end{center}
\end{figure}

Monolayer graphene consists of carbon atoms arranged in a two-dimensional honeycomb crystal structure, as shown in Fig. \ref{lattice}. The honeycomb structure consists of the triangular Bravais lattice with a basis of two atoms, labeled A and B. All the atoms from the same sublattice can be reached with primitive lattice vectors
\begin{equation} 
\label{latvec:list} 
\begin{split}
\mbf{a_1} &=  (\frac{1}{2},\frac{\sqrt{3}}{2})a,~~~~\text{and}  \\
\mbf{a_2} &=  (\frac{1}{2},-\frac{\sqrt{3}}{2})a,
\end{split} 
\end{equation}
where $a$ is the lattice constant and for graphene its value is $a = 2.46$ \AA.

The first Brillouin zone defined by these vectors is shown in Fig. \ref{lattice}(b). Of particular interest in the physics of graphene are the six corners of the first Brillouin zone which consist of three pairs of inequivalent points $\mbf{K}$ and $\mbf{K'}$. The position of these points in momentum space is given by
\begin{equation}
\label{ekkp1}
\mbf{K} = \frac{4\pi}{3\sqrt{3}a_{cc}}(1, 0) \qquad
\text{and} \qquad
\mbf{K'} = \frac{4\pi}{3\sqrt{3}a_{cc}}(-1, 0).
\end{equation}
The remaining corners can be connected to one of these points via translation by a reciprocal lattice vector. Note that the $\mbf{K}$ and $\mbf{K'}$ points are not related to the different sublattices of the graphene lattice but are a consequence of the two-dimensional hexagonal lattice structure.

Due to its strong covalent bonds, graphene exhibits excellent mechanical properties. Enormous values of Young's modulus and intrinsic strength of a defect-free graphene sheet were reported in Ref. \cite{cgra_mp01}. In this paper, Lee \textit{et al.} measured the elastic properties and intrinsic breaking strength of free-standing monolayer graphene membranes through nanoindentation by an atomic force microscope (AFM). They found a Young modulus of $E \approx 1$ TPa. This value is very close to the theoretical value E = 1.05 TPa \cite{radd_01}. Experiment further showed that the brittle fracture of graphene occurs at a critical stress equal to its intrinsic strength of $\sigma_{int} = 130$ GPa. This is the highest value ever measured for a material.
\section{Strain engineering}
\label{sec:1.0}
To describe graphene one can use the standard nearest-neighbor tight-binding Hamiltonian given by
\begin{equation}
\label{ch8e0}
H = \sum_{i}\epsilon_i c_{i}^{\dagger}c_i + \sum_{i,j} t_{ij}c_{i}^{\dagger}c_j,
\end{equation}
where $c_{i}^{\dagger}$($c_i$) is the creation (annihilation) operator for an electron at site $i$, $\epsilon_i$ is the onsite potential at site $i$, and $t_{ij}$ is the hopping energy between sites $i$ and $j$. Stretching graphene results in changes of the bond length between neighboring atoms in its lattice. The new positions of the carbon atoms are given by $\mbf{r}_i = \mbf{r}_i^{eq} + \mbf{u}$, where $\mbf{r}_i^{eq}$ is the equilibrium position of atom $i$ and $\mbf{u}$ is the displacement field. This change results in a modification of the hopping energy given by
\begin{equation}
\label{ch8e1}
t_{ij} = t_0 e^{-\beta(d_{ij}/a_0 - 1)},
\end{equation}
where $t_0 = 2.8$ eV is the equilibrium hopping energy, $a_0=0.142$ nm is the length of the unstrained $C-C$ bond, and $d_{ij}$ is the length of the strained bond between atoms $i$ and $j$. The decay factor $\beta = \partial \log t / \partial \log a \mid_{a=a_0} \approx 3.37$ \cite{rstm03} is the strained hopping energy modulation factor and has been extracted from experiment. As a check, one can calculate the energy of the next-nearest neighbour in graphene which gives $t' = 0.23$ eV and agrees well with estimates obtained from other techniques \cite{cll01, rstm04}. On the other hand, $d_{ij}$ can be calculated from the strain tensor as
\begin{equation}
\label{estx01}
d_{ij} = \frac{1}{a_0} \mbf{r}_{ij} \mbs{\varepsilon} \mbf{r}_{ij} = \frac{1}{a_0}\left( a_0^2 + \varepsilon_{xx}x_{ij}^2 + \varepsilon_{yy}y_{ij}^2  +2 \varepsilon_{xy}x_{ij}y_{ij} \right),
\end{equation}
where $\mbs{\varepsilon}$ is the strain tensor obtained from classical continuum mechanics and given by \cite{rstm05}
\begin{equation}
\label{estx02}
\varepsilon_{ij} = \frac{1}{2} \left( \partial_j u_i \partial_i u_j + (\partial_iu_z)(\partial_ju_z) \right),~~~~~ i, j = x, y.
\end{equation}
The spatial variation of the hopping energy is equivalent to the generation of a magnetic vector potential, $\mathbf{A} = (A_x, A_y, 0)$, which can be evaluated around the $\mathbf{K}$ point using\cite{ret1}
\begin{equation}
\label{ch8e2}
A_x - \mathtt{i} A_y = -\frac{1}{ev_F}\sum_j \delta t_{ij} e^{\mathtt{i}\mathbf{K}\cdot \mathbf{r_{ij}}},
\end{equation}
where the sum runs over all neighboring atoms of atom $i$, $v_F$ is the Fermi velocity, $\delta t_{ij} = (t_{ij} - t_0)$, and $\mathbf{r_{ij}} = \mathbf{r_i} - \mathbf{r_j}$.
We can obtain $\delta t_{ij}$ by expanding Eq. \eqref{ch8e1} to e.g. linear order
\begin{equation}
\label{ch8e21}
t_{ij} = t_0 \left( 1 - \frac{\beta}{a_0}(d_{ij} - a_0) \right),
\end{equation}
which using Eq. \eqref{estx01} transforms into
\begin{equation}
\label{ch8e22}
t_{ij} = t_0 \left( 1 - \frac{\beta}{a_0^2} \mbs{\delta}_i \mbs{\varepsilon} \mbs{\delta}_j \right).
\end{equation}
Inserting Eq. \eqref{ch8e22} into Eq. \eqref{ch8e2} one can show that the vector potential has the following form \cite{radd_11}
\begin{equation}
\label{ch8e23}
\mbf{A} = -\frac{\hbar \beta}{2ea_{cc}} \begin{pmatrix}
\varepsilon_{xx} - \varepsilon_{yy} \\ -2\varepsilon_{xy}
\end{pmatrix}.
\end{equation}
Eq. \eqref{ch8e23} leads to an effective low-energy Hamiltonian ($\mbf{k}= \mbf{K} + \mbf{q}$ and  $\left|\mbf{q}\right| \rightarrow 0$) given by
\begin{equation}
\label{ch8e24}
H_{\pm \mbf{K}}(\mbf{q}) = v_f \mbs{\sigma} (\mbf{q} \pm e\mbf{A}).
\end{equation}
The pseudo-magnetic field (PMF) is then obtained as
\begin{equation}
\label{ch8e3}
\mathbf{B_{ps}} = \mathbf{\bigtriangledown} \times \mathbf{A} = \left(0,0,\partial_x A_y - \partial_y A_x \right)= \left(0,0,B_{ps}\right).
\end{equation}
Here we use the subscript $"ps"$ to differentiate between the pseudo-magnetic field generated by strain from the applied external magnetic field. It is important to mention that the PMF calculated for the  $\textbf{K'}$ point has the opposite direction compared to the one in the $\textbf{K}$ point (see Eqs. \eqref{ch8e24}).

\begin{figure}[htbp]
\begin{center}
\includegraphics[width=12cm]{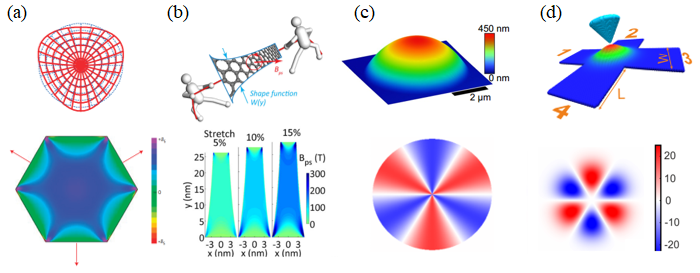}
\caption{(a - d) Different strain configurations (top panel) with the corresponding PMF profiles (bottom panel). Figures taken from Refs. \cite{rsq01, rsq02, rsq03, rsq04}.}
\label{f2}
\end{center}
\end{figure}

It is obvious from Eq. \eqref{ch8e2} that different strain configurations result into different profiles of the pseudo-magnetic field. This is illustrated in Fig. \ref{f2} where we show a few PMF profiles (bottom panel) generated by displacement fields shown in the top panel of the same figure. Guinea \textit{et al.} showed that a triaxial (triagonal) stretch results into a quasi-uniform PMF \cite{rsq01} (See Fig. \ref{f2}(a)). A similar type of field is generated if one bends a graphene nanoribbon as in Ref. \cite{rsq01a}. Unfortunately, uniaxial stretch does not produce a pseudo-magnetic field. However, Zhu \textit{et al.} showed that in the case of graphene ribbons of a non-uniform width uniaxial stretch will result in a quasi-uniform PMF profile \cite{rsq02}. This is shown in Fig. \ref{f2}(b). The first experimental observation of the generation of Landau levels by a pseudo-magnetic field in graphene was reported in Ref. \cite{ref_a01}. In their experiment, authors grew graphene on a Pt(111) substrate and found that a few triangular bubbles appear where graphene is stretched. In the strained region, Levy \textit{et al.} measured $dI/dV$ curves that exhibit well-defined Landau levels. From the Landau levels spacing, authors were able to extract a value of the magnetic field which was higher than 300 T. Obtaining a real magnetic field of such a high intensity in a laboratory is probably an impossible task.

On the other hand, most of the out-of-plane deformations, e.g. bumps and bubbles, result in a non-uniform distribution of the PMF. Numerous studies \cite{rsq03, rsq04, rsq05, rsq06, rsq07, rsq08, rsq09} showed that these types of strain result in alternating regions of positive and negative magnetic field, as shown in Figs. \ref{f2}(c-d). Of course, all the pseudo-magnetic field profiles shown in Fig. \ref{f2} are calculate for one  valley (at $\mathbf{K}$ point). The PMF in the other valley (at $\mathbf{K'}$ point), is similar to the one shown in Fig. \ref{f2} with the difference that positive and negative regions are now switched. Hence, half of the electrons feel magnetic field in one direction and the other half feel a magnetic field of the same strength but with opposite sign. Thus, the total average magnetic field in the whole structure is zero.

\begin{figure}[htbp]
\begin{center}
\includegraphics[width=11cm]{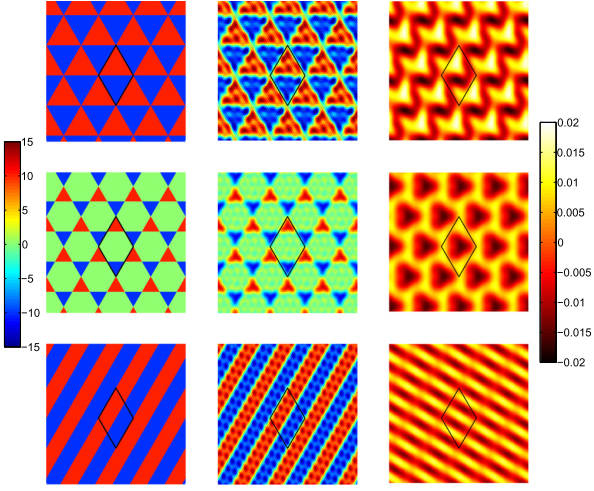}
\caption{Patterning a substrate can be used to engineer various PMF profiles. Left panel:  targeted PMF; middle panel: calculated PMF profile; right panel: required substrate topology. Left color scale: PMF value in Tesla. Right color scale: out-of-plane deformation in units of $10$ nm. Figure taken from Ref. \cite{rsq10}.}
\label{f3}
\end{center}
\end{figure}

In a recent paper by Jones and Pereira, a numerical recipe was given how to design a particular PMF profile by patterning the substrate on which a graphene layer is deposited \cite{rsq10}. When external pressure is applied, graphene stretches following the topology of the substrate. The presented method relies on a PDE (partial differential equations) - constrained optimization strategy to minimize the generic objective function which penalizes significant deviations between the induced and target PMFs \cite{rsq10}. In Fig. \ref{f3} we show the results of their calculations. Left column shows targeted PMF. Middle column shows the field profile calculated from the deformation field shown in the right column. Recently, those authors and co-workers used the same technique to optimize a graphene Corbino device for valley filltering applications \cite{rsq11}.
\section{Valleytronics}
\label{sec:2.0}
After the discovery of graphene, one of the major tasks was to use this material for electronic circuitry. Low dimensionality, high conductivity, and electron mobility makes this material ideal for high-frequency logic operations. Unfortunately, graphene is a semimetal. It has no band gap which severely limits its application in electronic industry. Hence, many groups tried to find a way to overcome this problem by opening a gap in the graphene spectrum. One of the most popular routes to do this is by enhancing the spin-orbit interaction in graphene \cite{rsoi01}. Another route envolves using graphene for spintronics. Exploiting the spin degree of freedom proved to be a very challenging task. Graphene has extremely long spin diffusion length (of order of few micrometers at room temperature \cite{rsoi02, rsoi03, rsoi04}) making this material suitable as a spin channel material. However, graphene is diamagnetic which means that in order to have spin-polarized current one would need either to inject it from a ferromagnetic leads or to induce spin ordering in graphene using e.g. ferromagnetic defects. Both approaches have a number of limitations \cite{rsoi05}. In the former case, injection efficiency is strongly limited by the conduction mismatch between graphene and a ferromagnetic lead \cite{rsoi05a}. In the latter case, transport properties of graphene would be highly affected by the presence and density of adatoms and/or vacancies.
\begin{figure}[htbp]
\begin{center}
\includegraphics[width=12cm]{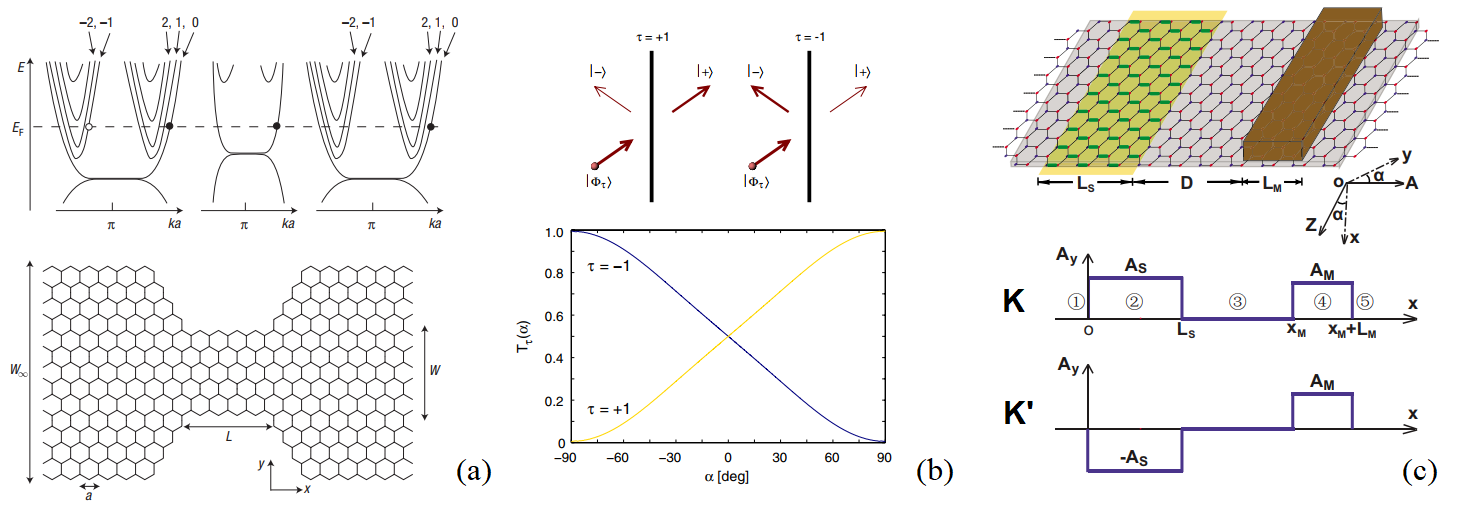}
\caption{(a) Energy spectrum of a graphene constriction. By tuning the Fermi level inside the constriction one is able to create valley polarized current. Figure taken from Ref. \cite{rvf02}. (b) (Top) The sublattice symmetric $\left| + \right>$ and antisymmetric $\left| - \right>$ components of the incident state $\left| \Phi_{\tau} \right>$ are transmitted and reflected, respectively. The thickness of each arrow indicates the probability the electron will follow the respective path. (Bottom) The probability that an incident electron at the Fermi level with valley index $\tau$ and angle of
incidence $\alpha$ will transmit through the line defect. Figure taken from Ref. \cite{r_add_1}. (c) Graphene nanoribbon with a stretched region (yellow) and a deposited ferromagnetic stripe (brown region). Bottom plot shows the profile of the gauge field along the ribbon for electrons from two valleys. Figure taken from Ref. \cite{r_add_31}.}
\label{f4a}
\end{center}
\end{figure}

Opening a band gap in graphene and exploiting its spin degree of freedom are both very challenging tasks. Efficiency of these devices would be very low and the implentation of the above mentioned proposals require extremely complex device geometries. However, graphene has one additional degree of freedom that we didn't make use of - two valleys. Recently, there has been a lot of interest to use the valley degree of freedom to encode information. The idea is similar as for spintronics where the spin degree of freedom is used as a bit of information, however, valleytronic devices would recquire less complex device designs. Using two valleys to encode information was first proposed in Ref. \cite{rvf01} for the AlAs quantum point contact (QPC) structure. In this study authors were able to create valley polarized currents due to the fact that the lowest energy levels of a QPC are occupied by heavy mass states along the QPC lateral dimension. Hence, depending on the orientation of the QPC one is able to filter either electons from the Y or X valley. 

The first proposal to use graphene for valleytronics was given in Ref. \cite{rvf02} where Rycerz \textit{et al.} showed that graphene nano-constrictions can be used to create valley polarized currents. The idea relies on the fact that the lowest energy sub-band in the case of a zigzag graphene nanoribbon is valley polarized. Thus, if one tunes the Fermi level in such a way that in the region of the constriction only this sub-band is occupied electrons from different valleys would propagate in opposite direction and consequently, current becomes valley polarized (See Fig. \ref{f4a}(a)). By adding a gate in the constriction region one can move the Fermi region to the valence band and even completely block the current flow (valley valve). This was later on explained as a consequence of parity of the lowest mode under the switch of the sublattice, i.e. the incident and transmitted modes have opposite parity for even number of atoms across the ribbon, leading to a complete reflection, while they have the same parity for odd number of atoms, leading to a complete transmission \cite{rvf03}. 

Since then, there have been numerous different proposals to use graphene for valleytronics applications. Several of them rely on the usage of line defects in graphene. In these systems, electron transmission through the defect depends on the valley degree of freedom \cite{r_add_1, r_add_11, r_add_111,r_add_1111}. Electrons approaching the line defect are reflected or transmitted by it depending on the angle of incidence (See top part of Fig. \ref{f4a}(b)). Hence, for some angles transmission of electrons from one valley is completely blocked while the electrons from the opposite valley transmit without loss, as shown in the bottom part of Fig. \ref{f4a}(b). Other proposals are based on  the addition of the specific mass term to the Hamiltonian that controls the valley isospin \cite{r_add_2, r_add_21}. Combination of strain and an external magnetic field is also a very popular choice to obtain a valley filter \cite{r_add_3, r_add_31, r_add_32}. Here, the pseudo-gauge field is used to break valley degeneracy and the external magnetic field to filter electrons from one valley (See Fig. \ref{f4a}(c)). In the rest of this section, we discuss the most interesting valley filters that rely solely on the use of strain for the generation of valley polarized currents and the first experimental realizations of valley filters.
\begin{figure}[htbp]
\begin{center}
\includegraphics[width=12.cm]{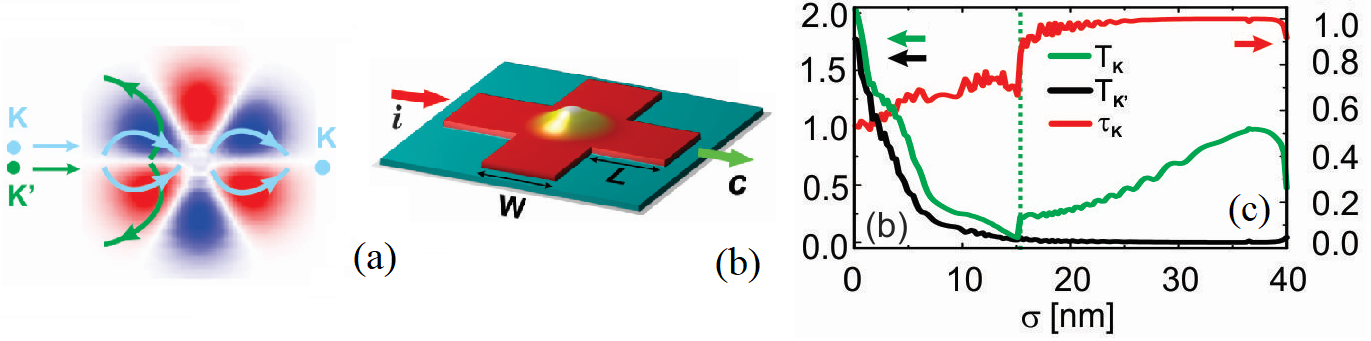}
\caption{(a) Sketch of flow of electrons from different valleys through the strained region. (b) Schematics of a cross shaped Hall bar with a bump in the center. (c) Transmission probabilities $T_K$, $T_{K'}$, $\tau_K$ versus the width of the bump, $\sigma$. Calculations are performed for $E_F = 0.1$ eV and $h_0 = 20$ nm. Figure taken from Ref. \cite{rvf05}.}
\label{f4}
\end{center}
\end{figure}

Controlling the edges of the sample is an extremely difficult task. Easier route to separate electrons from different valleys is by the use of strain. As mentioned in Sec. \ref{sec:1.0}, straining the graphene lattice results in the appearance of a PMF. This field has opposite direction for electrons that originate from different valleys (See Eqs. \eqref{ch8e2} and \eqref{ch8e3}). We can use this fact to separate electrons from the two valleys. The idea is sketched in Fig. \ref{f4}(a). Current carried by both, $\mbf{K}$ and $\mbf{K'}$ electrons, flows towards a bump region. Here, electrons from one valley, e.g. $\mbf{K}$, feel the magnetic field as plotted in Fig. \ref{f4}(a) and move along the zero-PMF line towards the other end of the bump. Electrons from  $\mbf{K'}$ valley, feel a magnetic field of opposite sign and consequently move in opposite direction, i.e. they are reflected by the bump. Hence, on the opposite side of the bump a valley-polarized current is generated. We tested this on a four-terminal structure with a bump in the middle, as shown in Fig. \ref{f4}(b). The bump has a Gaussian-like profile given by $z(\mbf{r})= h_0 exp(-\mbf{r}^2/2\sigma^2)$. Current is injected at lead $i$ and collected at the opposite side of the bump, lead $c$. The two perpendicular leads are added to collect any current reflected from the bump. The results of our calculations are shown in Fig. \ref{f4}(c). Here we show how the transmission of electrons from $\mbf{K}$, $\mbf{K'}$, as well as the polarization $t_K$ ($t_K = T_K/(T_K + T_{K'})$), change with the width of the bump. One notices that as $\sigma$ increases the total transmission first goes down which simply means that the current is heavily scattered by the bump. However, after a certain value of $\sigma$, transmission probability $T_K$ starts to increase while $T_{K'}$ stays low. In this regime the bump works as a valley filter. Conducting channels inside the bump open and valley polarized current is generated at the opposite side of the bump \cite{rvf05}. 

\begin{figure}[htbp]
\begin{center}
\includegraphics[width=12.cm]{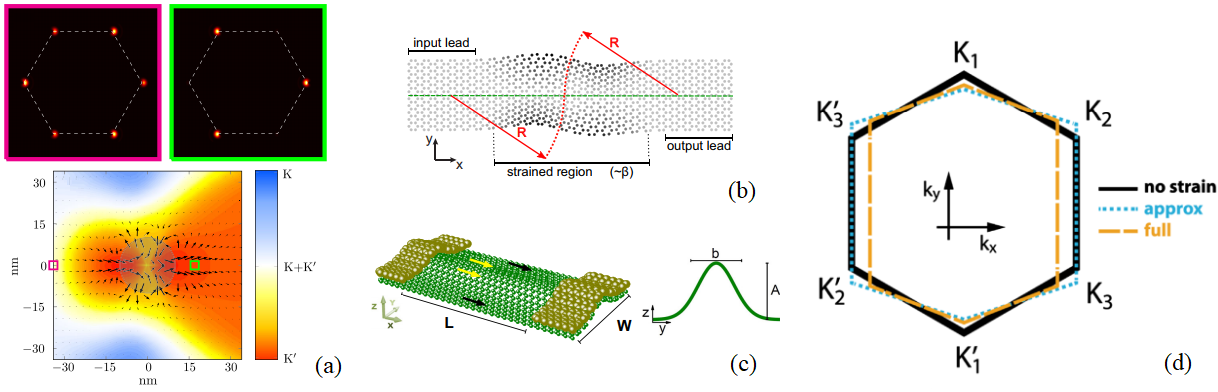}
\caption{(a) Top panel shows the $k$-space occupation of the magenta and green square from the bottom panel. Bottom panel shows local current map in the sample. Figure taken from Ref. \cite{rvf06}. (b) Sketch of the strained graphene ribbon. Figure taken from Ref. \cite{ref_a02}. (c) Graphene nanoribbon with a Gaussian fold. Figure taken from Ref. \cite{rvf07}. (d) Change of the first Brillouin zone of graphene with strain. The black, blue, and orange line show the unstrained case, using linear approximation of Eq. \eqref{ch8e2}, and using Eq. \eqref{ch8e2}, respectively. Figure taken from Ref. \cite{rsq07}.}
\label{f4b}
\end{center}
\end{figure}

Similar observations were made in Ref. \cite{rvf06}. In this paper authors calculated the spectral density for each $\mbf{k}$ value at the two opposite sides of the Gaussian bump for an infinitely extending graphene sheet. Fourier maps showed contributions from both valleys for the region before the bubble while in the region after the bubble only electrons from a single valley were detected, as shown in Fig. \ref{f4b}(a).  Furthermore, upon a change of the sign of the Fermi energy one can switch between the two valleys in the filtered region.

Cavalcante \textit{et al.} studied snake states transport in a graphene nanoribbon with a  bended region as in Ref. \cite{rsq01}. This type of strain, shown in Fig. \ref{f4b}(b), generates a (pseudo-) magnetic barrier which allows snake state transport along the length of the ribbon. In their study (Ref. \cite{ref_a02}), authors use the fact that electrons from opposite valleys propagate in opposite direction to generate valley dependent snake state transport. The efficiency of this valley filter can be improved by carefully tuning the Fermi energy.

In Ref. \cite{rvf07} Carillo-Bastos \textit{et al.} used a graphene nanoribbon with a Gaussian fold, as shown in Fig. \ref{f4b}(c). Enhanced LDOS around the fold region was noticed as a consequence of the sublattice symmetry breaking. LDOS plots revealed a current split in this region. Namely, part of the current along the center of the strained fold was coming from states from one valley. The region around sides of the fold, on the other hand, is filled with states from the other valley. Thus, using a fold-like structure one is able to spatially separate electrons from different valleys. If edge disorder is present in the structure, the edge states as well as states further away from the center of the ribbon would be destroyed. In this way, one is able to generate valley polarized currents at a given point \cite{rvf07}.

The results of Rycerz \textit{et al.} were recently confirmed experimentally in Ref. \cite{rvf04} where authors used a graphene ribbon with a non-uniform width (in)conjunction with a p-n junction. The p-n interface can be moved along the length of the ribbon by means of electrostatic gating. Authors observed that the conductance changed as the p-n interface crosses the point where the width of the junction change. They argue that this is due to the change of the isospin configuration of the edges. Namely, if the bottom and the top edge of the ribbon have the same isospin, reflection is forbidden. This will change as the p-n interface is moved along the ribbon and a drop in the conductance will occur. A set of oscillations (of order of $e^2/h$) appear in the conductance plot which are sensitive to the position of the p-n interface but do not change with doping or magnetic field. This suggests that these conductance oscillations are the first experimental confirmation of the effect of the local isospin configuration of the edges on transport in graphene.

Recently, Georgi \textit{at al.} observed sublattice polatization in a strained graphene sample \cite{rvf08}. Authors used an STM tip to locally deform the graphene lattice and at the same time map changes in the LDOS. By carefully tuning the strain they were able to observe up to 50 $\%$ difference in the LDOS at different sublattices. Although, sublattice polarization itself does not directly imply valley polarization, results of Georgi \textit{at al.} showed that the PMF generated by the STM tip indeed shows regions with positive and negative pseudo-magnetic field. Furthermore, the strength of this field is large enough to cause valley dependent deflection of electrons. Hence, this result represents an important advance towards the experimental realization of valley filters.
\section{Strain sensing}
\label{sec:3.0}
Straining graphene locally results in a change of the global electronic properties of the system, e.g. resistance changes. This is due to the fact that the $\mbf{K}$ point changes position \cite{rss01} in the strained region resulting in electron scattering due to the mismatch of the Dirac points in the strained and unstrained region (See Fig. \ref{f4b}(d)). The efficiency of electron scattering is proportional to the amount of applied strain. This provides the possibility to use a resistance measurement to calculate the amount of strain. Furthermore, if straining of the graphene sheet is a consequence of external pressure we can use such a resistance measurement to get information about the pressure, i.e. realize a pressure sensor.

Graphene membranes are ideal candidates for pressure sensing applications due to their large stretchability and their electrical conducting properties. Graphene is an impermeable membrane for almost all standard gases including helium \cite{rss02, rss03, ref_a03}. It clamps firmly to almost all substrates achieving extremely large adhesion energies \cite{rsq03}. Kleshtanova \textit{et al.} showed that graphene membranes with small radius (less than 10 nm) can sustain up to 1 GPa of pressure without rapture \cite{rss04}. This makes graphene a suitable candidate for extremely high pressure sensing applications.

\begin{figure}[htbp]
\begin{center}
\includegraphics[width=12.cm]{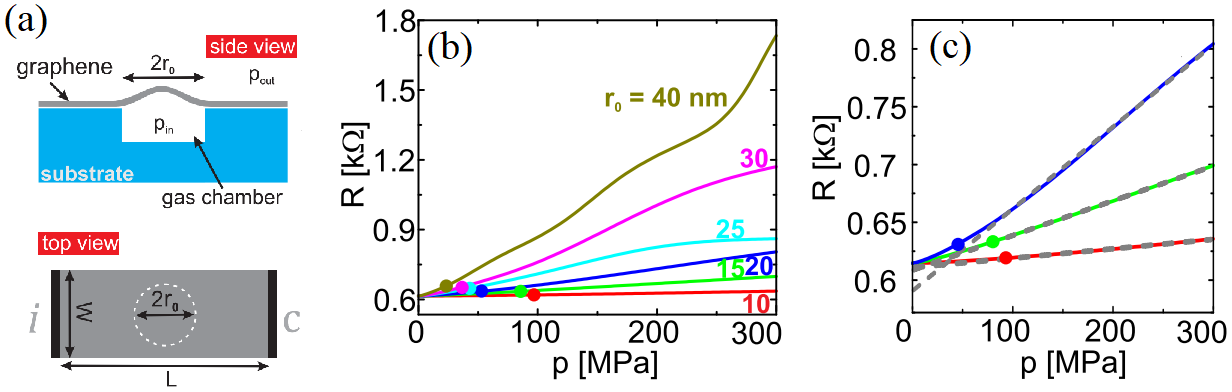}
\caption{(a) Schematics of a graphene pressure sensor. (b) Resistance versus applied pressure for different values of the bubble radius, $r_0$. Circles show the pressure at which the bubble height reaches 10$\%$ of its radius (membrane model is not valid below this value \cite{rss05}). (c) The resistances for the three smallest bubbles from (b) together with the corresponding linear fits (gray dashed lines). Figure taken from Ref. \cite{rss06}.}
\label{f5}
\end{center}
\end{figure}

To test this we calculated the resistance of a graphene membrane shown in Fig. \ref{f5}(a). The system consists of a substrate with a small chamber of radius $r_0$ etched in it. Over the substrate a graphene layer is deposited. Thus, above the etched part the graphene sheet is not supported by a substrate, i.e. behaves as a free-standing membrane of radius $r_0$. The difference between the pressure inside and outside the chamber causes a bending/bulging of the graphene membrane. This will generate a pseudo-magnetic field and electrons will be scattered by the strained region. All relevant formulas regarding the problem of straining circular graphene membranes are given in Ref. \cite{rss05}. We assume that due to the strong adhesion energy graphene clamps firmly to the substrate, hence, the bending stiffness of graphene can be neglected. The resulting PMF has a three-fold symmetry with alternating positive and negative regions similar to the case of Gaussian-like bumps (See Figs. \ref{f2}(c-d) and \ref{f4}(a)). 

The results of our calculations are shown in Fig. \ref{f5}(b). Here we used $L = 2W = 200$ nm and $E_F = 0.2$ eV. Notice that for membranes of small radius the resistance changes linearly with pressure. This is clearly shown in Fig. \ref{f5}(c) where we plot the resistance of the three smallest bubbles with their corresponding linear fits (gray dashed lines). Sensitivity, on the other hand, increases with the size of the bubble. For the largest bubble, $r_0 = 40$ nm, resistance increases 3 times when pressure of 300 MPa is applied. The curve shows a highly non-linear trend versus the applied pressure. This is explained by the fact that in the case of large bubbles, interference between different propagating modes occurs which differs from the case of small bubbles where only a few modes are affected by the bubble \cite{rss06}. The calculations of the resistance were repeated for a few graphene stripes of different widths (in the range from 80 nm to 200 nm), and the results always showed a linear dependence on the pressure for bubbles with $r_0 < 0.15W$. This opens the possibility of scaling our results due to the fact that the ratio $r_0=W$ is the important quantity for defining the behavior of the resistance \cite{rss06}. Efficiency of the sensor is characterized by the gauge factor given by
\begin{equation}
\label{ess01}
GF = \frac{\Delta R/R}{\varepsilon},
\end{equation}
where $\Delta R/R$ is the relative change of the resistance with the strain and $\varepsilon$ is the maximal strain in the structure. Our simulations reveal that for $r_0 = 20$ nm we obtain $GF \approx 8$, while for $r_0 = 30$ nm we have $GF \approx 18$. However, linearity is lost in the latter case.

In Ref. \cite{rss07} authors measured the resistance of a suspended graphene nanoribbon in a nanoindentation experiment. Here, they used a wedge indentation tip to uniaxially strain a graphene ribbon and at the same time measure the change in the resistance. Measured Young's modulus was of the order of 335 N/m, which is in good agreement with previous experiments, and a gauge factor of $GF=1.9$ was obtained. The low gauge factor can be due to the fact that uniaxial stretch does not generate a psuedo-magnetic field. Similar results were obtained in Ref. \cite{rss08} with $GF=1.6$. On the other hand, Lee \textit{et al.} measured a gauge factor of 6.1 in their device made of printed graphene films \cite{rss09}. They observe a resistance increase from $\approx 492$ to $\approx 522$ k$\Omega$ when the graphene film was strained up to 1$\%$. Experiment showed that the resistance change is reproducible even after hundreds of repetitions. This value of gauge factor is higher than for conventional strain gauges based on metal alloys and thus shows a great potential for application of graphene in strain sensing.
\section{Acknowledgment}
This work was supported by the Flemish Science Foundation (FWO-Vl).

% BibTeX users please use
% \bibliographystyle{}
% \bibliography{}
%
% Non-BibTeX users please follow the syntax
% the syntax of "referenc.tex" for your own citations
%%%%%%%%%%%%%%%%%%%%%%%% referenc.tex %%%%%%%%%%%%%%%%%%%%%%%%%%%%%%
% sample references
% "physics"
%
% Use this file as a template for your own input.
%
%%%%%%%%%%%%%%%%%%%%%%%% Springer-Verlag %%%%%%%%%%%%%%%%%%%%%%%%%%

%
% BibTeX users please use
% \bibliographystyle{}
% \bibliography{}

\begin{thebibliography}{99.}
% Use the following syntax and markup for your references

% Book, authored
%\bibitem{monograph} H. Ibach, H. L\"uth, \textit{Solid-State
%Physics}, 2nd edn. (Springer, Berlin Heidelberg New York, 1996),
%pp.\,45--56

% Journal article
\bibitem{rgr1}K. S. Novoselov, A. K. Geim, S. V. Morozov, D. Jiang, Y. Zhang, S. V. Dubonos, I. V. Grigorieva, and A. A. Firsov, Science \textbf{306}, 666 (2004)
\bibitem{r_add_4} N. D. Mermin and H. Wagner, Phys. Rev. Lett. \textbf{17}, 1133 (1966)
\bibitem{r45} M. Takagi, J. Phys. Soc. Jpn. \textbf{9}, 359 (1954)
\bibitem{r46} J. A. Venables, G. D. T. Spiller, and M. Hanb\"{u}cken, Rep. Prog. Phys.
\textbf{47}, 399 (1984)
\bibitem{r48} E. Gibney, Nature \textbf{522}, 274 (2015)
\bibitem{rthes} S. P. Milovanovic, Electronic transport properties in nano- and micro-engineered graphene structures, University of Antwerp (2017) \\
web: \texttt{nano.uantwerpen.be/cmt/PhD$\_$slavisa$\_$milovanovic.pdf} 
\bibitem{cgra_mp01} C. Lee, X. Wei, J. W. Kysar, and J. Hone, Science \textbf{321}, 385 (2008)
\bibitem{radd_01} F. Liu, P. Ming, and J. Li, Phys. Rev. B \textbf{76}, 064120 (2007).
\bibitem{rstm03} V. M. Pereira, A. H. Castro Neto, and N. M. R. Peres, Phys. Rev. B \textbf{80}, 045401 (2009)
\bibitem{cll01} A. H. Castro Neto, F. Guinea, N. M. R. Peres, K. S. Novoselov, and
A. K. Geim, Rev. Mod. Phys. \textbf{81}, 109 (2009)
\bibitem{rstm04} B. Partoens and F. M. Peeters, Phys. Rev. B \textbf{74}, 075404 (2006)
\bibitem{rstm05}  L. D. Landau and E. M. Lifshitz, Theory of elasticity - Course of Theoretical Physics: volume 7, (Springer, Berlin 1986)
\bibitem{ret1} V. M. Pereira and A. H. Castro Neto, Phys. Rev. Lett. \textbf{103}, 046801 (2009)
\bibitem{radd_11} M. Vozmediano, M. Katsnelson, and F. Guinea, Phys. Reports \textbf{496}, 109 (2010)
\bibitem{rsq01}  F. Guinea, M. I. Katsnelson, and A. K. Geim, Nat. Phys. \textbf{6}, 30 (2009)
\bibitem{rsq01a} F. Guinea, A. K. Geim, M. I. Katsnelson, and K. S. Novoselov, Phys. Rev. B \textbf{81}, 035408 (2010)
\bibitem{rsq02} S. Zhu, J. A. Stroscio, and T. Li, Phys. Rev. Lett. \textbf{115}, 245501 (2015)
\bibitem{ref_a01} N. Levy, S. A. Burke, K. L. Meaker, M. Panlasigui, A. Zettl1, F. Guinea, A. H. Castro Neto, and M. F. Crommie, Science \textbf{329}, 544 (2010)
\bibitem{rsq03} S. P. Koenig, L. Wang, J. Pellegrino, and J. S. Bunch, Nat. Nanotech. \textbf{7}, 728 (2012)
\bibitem{rsq04} S. P. Milovanovic and F. M. Peeters, J. Phys: Condens. Matter \textbf{29}, 075601 (2016)
\bibitem{rsq05} K.-J. Kim, Ya. M. Blanter, and Kang-Hun Ahn, Phys. Rev. B \textbf{84} 081401(R) (2011)
\bibitem{rsq09} F. de Juan, A. Cortijo, M. A. H. Vozmediano, and A. Cano, Nat. Phys.
\textbf{7}, 810 (2011)
\bibitem{rsq06} N. N. Klimov, S. Jung, S. Zhu, T. Li, C. A. Wright, S. D. Solares, D. B. Newell, N. B. Zhitenev, and J. A. Stroscio, Science \textbf{336}, 1557 (2012)
\bibitem{rsq07} M. R. Masir, D. Moldovan, and F.M. Peeters, Solid State Comm. \textbf{175-176}, 76 (2013)
\bibitem{rsq08} M. Settnes, S. R. Power, and A.-P. Jauho, Phys. Rev. B \textbf{93}, 035456
(2016)
\bibitem{rsq10} G. W. Jones and V. M. Pereira, New J. Phys. \textbf{16}, 093044 (2014)
\bibitem{rsq11} G. W. Jones, D. A. Bahamon, A. H. Castro Neto, and V. M. Pereira, Nano Lett. \textbf{17}, 5304 (2017)
\bibitem{rsoi01} C. L. Kane and E. J. Mele, Phys. Rev. Lett. \textbf{95}, 226801 (2005)
\bibitem{rsoi02} N. Tombros, C. Jozsa, M. Popinciuc, H. T. Jonkman, and B. J. van Wees, Nature \textbf{448}, 571 (2007)
\bibitem{rsoi03} T.-Y. Yang, J. Balakrishnan, F. Volmer, A. Avsar, M. Jaiswal, J. Samm, S. R. Ali, A. Pachoud, M. Zeng, M. Popinciuc, G. G\"{u}ntherodt, B. Beschoten, and B. \"{O}zyilmaz, Phys. Rev. Lett. \textbf{107}, 047206 (2011) 
\bibitem{rsoi04} B. Dlubak, M.-B. Martin, C. Deranlot, B. Servet, S. Xavier, R. Mattana,	M. Sprinkle, C. Berger,	W. A. De Heer, F. Petroff, A. Anane, P. Seneor, and A. Fert, Nat. Phys. \textbf{8}, 557 (2012)
\bibitem{rsoi05} W. Han, R. K. Kawakami, M. Gmitra, and J. Fabian, Nat. Nanotech. \textbf{9}, 794 (2014)
\bibitem{rsoi05a} G. Schmidt, D. Ferrand, L. W. Molenkamp, A. T. Filip, and B. J. van Wees, Phys. Rev. B \textbf{62}, R4790(R) (2000)
\bibitem{rvf01} O. Gunawan, B. Habib, E.P. De Poortere, and M. Shayegan, Phys. Rev. B \textbf{74}, 155436 (2006)
\bibitem{rvf02} A. Rycerz, J. Tworzyd\l{}o, and C. W. J. Beenakker, Nat. Phys. \textbf{3}, 172 (2007)
\bibitem{rvf03} A. R. Akhmerov, J. H. Bardarson, A. Rycerz, and C. W. J. Beenakker, Phys. Rev. B \textbf{77}, 205416 (2008)
\bibitem{r_add_1} D. Gunlycke and C. T. White, Phys. Rev. Lett. \textbf{106}, 136806 (2011)
\bibitem{r_add_11} Y. Liu, J. Song, Y. Li, Y. Liu, and Q. Sun, Phys. Rev. B \textbf{87}, 195445 (2013)
\bibitem{r_add_111} J.-H. Chen, G. Aut\'{e}s, N. Alem, F. Gargiulo, A. Gautam, M. Linck, C. Kisielowski, O. V. Yazyev, S. G. Louie, and A. Zettl, Phys. Rev. B \textbf{89}, 121407(R) (2014)
\bibitem{r_add_1111} V. H. Nguyen, P. Dollfus, and J.-C. Charlier, Phys. Rev. Lett. \textbf{117}, 247702 (2016)
\bibitem{r_add_2} D. Moldovan, M. Ramezani Masir, L. Covaci, and F. M. Peeters, Phys. Rev. B \textbf{86}, 115431 (2012)
\bibitem{r_add_21} M. Ramezani Masir, A. Matulis, and F. M. Peeters, Phys. Rev. B \textbf{84}, 245413 (2011)
\bibitem{r_add_3} Z. Wu, F. Zhai, F. M. Peeters, H. Q. Xu, and K. Chang, Phys. Rev. Lett. \textbf{106}, 176802 (2011)
\bibitem{r_add_31} F. Zhai, X. Zhao, K. Chang, and H. Q. Xu, Phys. Rev. B \textbf{82}, 115442 (2010)
\bibitem{r_add_32} T. Fujita, M. B. A. Jalil, and S. G. Tan,  Appl. Phys. Lett. \textbf{97}, 043508 (2010)
\bibitem{rvf05} S. P. Milovanovic and F. M. Peeters, Appl. Phys. Lett. \textbf{109}, 203108 (2016).
\bibitem{rvf06} M. Settnes, S. R. Power, M. Brandbyge, and A.-P. Jauho, Phys. Rev. Lett. \textbf{117}, 276801 (2016)
\bibitem{ref_a02} L. S. Cavalcante, A. Chaves, D. R. da Costa, G. A. Farias, and F. M. Peeters, Phys. Rev. B \textbf{94}, 075432 (2016)
\bibitem{rvf07} R. Carrillo-Bastos, C. Leon, D. Faria, A. Latg\'{e}, E. Y. Andrei, and N. Sandler, Phys. Rev. B \textbf{94}, 125422 (2016)
\bibitem{rvf04} C. Handschin, P. Makk, P. Rickhaus, R. Maurand, K. Watanabe, T. Taniguchi, K. Richter, M.-H. Liu, and C. Sch\"{o}nenberger, Nano Lett. \textbf{17}, 5389 (2017)
\bibitem{rvf08} A. Georgi, P. Nemes-Incze, R. Carrillo-Bastos, D. Faria, S. Viola Kusminskiy, D. Zhai, M. Schneider, D. Subramaniam, T. Mashoff, N. Michael Freitag, M. Liebmann, M. Pratzer, L. Wirtz, C. R. Woods, R. Vladislavovich Gorbachev, Y. Cao, K. S. Novoselov, N. Sandler, and M. Morgenstern, Nano Lett. \textbf{17}, 2240 (2017)
\bibitem{rss01} V. M. Pereira and A. H. Castro Neto, Phys. Rev. Lett. \textbf{103}, 046801 (2009)
\bibitem{rss02} O. Leenaerts, B. Partoens, and F. M. Peeters, Appl. Phys. Lett. \textbf{93}, 193107 (2008)
\bibitem{rss03} R. R. Nair, H. A. Wu, P. N. Jayaram, I. V. Grigorieva, and A. K. Geim,
Science \textbf{335}, 442 (2012)
\bibitem{ref_a03} J. Scott Bunch, Scott S. Verbridge, Jonathan S. Alden, Arend M. van der Zande, Jeevak M. Parpia, Harold G. Craighead and Paul L. McEuen, Nano Lett. \textbf{8}, 2458 (2008)
\bibitem{rss04} E. Khestanova, F. Guinea, L. Fumagalli, A. K. Geim, and I. V. Grigorieva,
Nat. Commun. \textbf{7}, 12587 (2016)
\bibitem{rss05} K. Yue, W. Gao, R. Huang, and K. M. Liechti, J. Appl. Phys. \textbf{112}, 083512 (2012)
\bibitem{rss06} S. P. Milovanovic, M. \v{Z}. Tadi\'{c}, and F. M. Peeters, Appl. Phys. Lett. \textbf{111}, 043101 (2017)
\bibitem{rss07} M. Huang, T. A. Pascal, H. Kim, W. A. Goddard, and J. R. Greer, Nano
Lett. \textbf{11}, 1241 (2011)
\bibitem{rss08} S.-E. Zhu, M. Krishna Ghatkesar, C. Zhang, and G. C. A. M. Janssen, Appl. Phys. Lett. \textbf{102}, 161904 (2013)
\bibitem{rss09} Y. Lee, S. Bae, H. Jang, S. Jang, S.-E. Zhu, S. H. Sim, Y. I. Song, B. H.
Hong, and J.-H. Ahn, Nano Lett. \textbf{10}, 490 (2010)






\end{thebibliography}

%%%%%%%%%%%%%%%%%%%%%%%%%%%%%%%%%%%%%%%%%%%%%%%%%%%%%%%%%%%%%%%%%%%%%%  }

%%%%%%%%%%%%%%%%%%%%%%%%%%%%%%%%%%%%%%%%%%%%%%%%%%%%%%%%%%%%%%%%%%%%%%

\printindex
\end{document}